\documentclass[conference]{IEEEtran}
\IEEEoverridecommandlockouts
\PassOptionsToPackage{bookmarks={false}}{hyperref} 
\usepackage[a-1b]{pdfx}

\usepackage{cite}
\usepackage{stfloats}
\usepackage{amsmath,amssymb,amsfonts}
\usepackage{adjustbox}
\usepackage{algorithmic}
\usepackage{graphicx}
\usepackage{textcomp}
\usepackage{xcolor}
\usepackage{caption}
\usepackage{soul}
\usepackage{subfig}
\def\BibTeX{{\rm B\kern-.05em{\sc i\kern-.025em b}\kern-.08em
    T\kern-.1667em\lower.7ex\hbox{E}\kern-.125emX}}
    
\begin{document}

\newcommand{\SubItem}[1]{
    {\setlength\itemindent{15pt} \item[-] #1}
}
\newcommand{\bh}[1]{\textcolor{blue}{Comment: #1}}



\title{\huge {Comprehensive RF Dataset Collection and Release: A Deep Learning-Based Device Fingerprinting Use Case}}


\author{\IEEEauthorblockN{Abdurrahman Elmaghbub and Bechir Hamdaoui}
\IEEEauthorblockA{\textit{School of Electrical Engineering and Computer Science},
\textit{Oregon State University}\\
\{elmaghba, hamdaoui\}@oregonstate.edu}

}

\maketitle

\begin{abstract}
Deep learning-based RF fingerprinting has recently been recognized as a potential solution for enabling newly emerging wireless network applications, such as spectrum access policy enforcement, automated network device authentication, and unauthorized network access monitoring and control.
Real, comprehensive RF datasets are now needed more than ever to enable the study, assessment, and validation of newly developed RF fingerprinting approaches. In this paper, we present and release a large-scale RF fingerprinting dataset, collected from $25$ different LoRa-enabled IoT transmitting devices using USRP B$210$ receivers. Our dataset consists of a large number of SigMF-compliant binary files representing the I/Q time-domain samples and their corresponding FFT-based files of LoRa transmissions. 
This dataset provides a comprehensive set of essential experimental scenarios, considering both indoor and outdoor environments and various network deployments and configurations, such as the distance between the transmitters and the receiver, the configuration of the considered LoRa modulation, the physical location of the conducted experiment, and the receiver hardware used for training and testing the neural network models.

\end{abstract}


\section{Introduction}

This paper presents and releases a comprehensive dataset consisting of massive RF signal data captured from $25$ LoRa-enabled transmitters using Ettus USRP B$210$ receivers. The RF dataset provides the research community with a set of first-of-its-kind real-world setups, all helpful in enabling the study of the effectiveness and robustness of deep learning-based wireless techniques, such as RF/device fingerprinting. These experimental setups cover and provide a comprehensive set of practical scenarios for indoor and outdoor environments while considering various realistic network deployment variability across time, location, hardware, and modulation configuration. These are obtained by varying network deployment and configuration parameters, such as the distance between the transmitters and the receiver, the configuration of the LoRa protocol, the physical location of the experiment, and the receiver hardware used for collecting the data samples and for training/testing the neural network models.

As research communities move away from model-based to data-driven solutions, many recently proposed frameworks on device/RF fingerprinting have also shifted from model-based classification approaches to deep learning-based approaches. Although recently proposed deep learning-based approaches have shown promising results, 
some still rely on synthesized data for evaluation and validation, and most of those that rely on real datasets are evaluated on general indoor and outdoor setups, leaving behind many questions that need to be answered before claiming the feasibility and superiority of their proposed techniques in real-world settings. 

Other essential experimental scenarios, which are very useful for assessing the capability and robustness of deep learning-based fingerprinting techniques, have not been investigated. These essential scenarios allow for the study of the impact of various deployment parameters on the performances achievable by these techniques, such as the impact of the distance between the transmitters and the receiver, the configuration of the used protocol, the physical location of the conducted measurements, and the hardware impairments of the receiver used in the training and inference stages. 

Another important scenario that is also missing is the study of the performance of deep learning models when the RF signals are captured during concurrent transmissions where some portions of the incoming packets overlap.  
 
A major challenge the wireless research community has been facing is the lack of comprehensive, publicly available datasets that could serve as benchmarks for the device/RF classification and fingerprinting techniques \cite{al2020exposing}. 
Having public and easily accessible dataset benchmarks has indeed been one of the main drivers for innovation and idea maturity in closely related fields, like image recognition and natural language processing, in addition to creating collaboration opportunities among researchers. Therefore, such efforts must be mimicked in the wireless community to be able to foster innovation in this domain as well.

 
 \subsection{What Distinguishes Our Dataset From Existing Ones?}
 
Researchers in \cite{aernouts2018sigfox} presented three LPWAN (Sigfox and LoRAWAN) datasets collected in an outdoor environment aimed at evaluating location fingerprinting algorithms. These datasets have been collected over three months in both rural and urban areas.
Sigfox-rural and Sigfox-urban datasets consist respectively of $25$k+ and $14$k+ Sigfox messages, whereas LoRaWAN dataset consists of $123$k+ LoRaWAN messages. These messages include a couple of protocol information, time of reception, and GPS location information but do not include device labels, and therefore, can not be used for supervised deep learning-based device classification. The closest existing work to our work is the recently released work at Northeastern University~\cite{al2020exposing}, which collected and released a massive dataset of IEEE 802.11 a/g (WiFi) standard data obtained from 20 wireless devices with identical RF circuitry over several days in (a) an anechoic chamber, (b) in-the-wild testbed, and (c) with cable connections. The focus of the Northeastern dataset is to explore the impact of the wireless channel on the performance of deep learning-based RF fingerprinting models. The dataset is limited in terms of the covered scenarios, and it is for WiFi signals only. Unfortunately, to the best of our knowledge, there are still no public datasets for LoRa device fingerprinting nor datasets that include the diverse experimental scenarios we mentioned. Our dataset presented in this paper fulfills the need for a large-scale LoRa dataset covering a wide range of diverse scenarios for a more comprehensive evaluation and validation.  
 
      \begin{table*}
\begin{adjustbox}{width=2\columnwidth, height = 0.17\columnwidth, center}


\begin{tabular}{|l|c|c|c|c|c|c|c|c|c|}
\hline
\textbf{Setups}                                    & {\color[HTML]{333333} \textbf{\begin{tabular}[c]{@{}c@{}}Number of\\ Transmitters\end{tabular}}} & { \textbf{\begin{tabular}[c]{@{}c@{}}Number of\\  Receivers\end{tabular}}} & \textbf{Protocol} & \textbf{\begin{tabular}[c]{@{}c@{}}Number \\ of Days\end{tabular}} & \textbf{\begin{tabular}[c]{@{}c@{}}Transmissions \\ per Device\end{tabular}} & \textbf{\begin{tabular}[c]{@{}c@{}}Duration per \\ Transmission\end{tabular}} & \textbf{Distances} & \textbf{Environment} & \textbf{Representation} \\ \hline
{1) Diff Days Indoor}                              & {\color[HTML]{333333} 25}                                                                        & {\color[HTML]{333333} 1}                                                                       & LoRa              & 5             & 10                                                                           & 20s                                                                           & 5m                 & Indoor               & IQ/FFT                  \\ \hline
{2) Diff Days Outdoor}                             & 25                                                                                               & 1                                                                                              & LoRa              & 5             & 10                                                                           & 20s                                                                           & 5m                 & Outdoor              & IQ/FFT                  \\ \hline
{3) Diff Days Wired}                               & 25                                                                                               & 1                                                                                              & LoRa              & 5             & 10                                                                           & 20s                                                                           & 5m                 & Wired                & IQ/FFT                  \\ \hline
{4) Diff Distances}                            & 25                                                                                               & 1                                                                                              & LoRa              & 1             & 4                                                                            & 20s                                                                           & 5,10,15,20m        & Outdoor              & IQ/FFT                  \\ \hline
\multicolumn{1}{|l|}{{5) Diff Configurations}} & 25                                                                                               & 1                                                                                              & LoRa              & 1             & 4                                                                            & 20s                                                                           & 5m                 & Indoor               & IQ/FFT                  \\ \hline
{6) Diff Locations}                            & 25                                                                                               & 1                                                                                              & LoRa              & 1             & 3                                                                            & 20s                                                                           & 5m                 & 2 Indoor, 1 Outdoor  & IQ/FFT                  \\ \hline
{7) Diff Receivers}                            & 25                                                                                               & 2                                                                                              & LoRa              & 1             & 2                                                                            & 20s                                                                           & 5m                 & Indoor               & IQ/FFT                  \\ \hline
\end{tabular}

\end{adjustbox}
\caption{Summary of Experimental Setups/Scenarios.}
\label{summary}
\end{table*}

\subsection{Our LoRa Dataset in Brief}
Our RF dataset provides both time-domain I/Q samples and corresponding FFT samples collected using an IoT testbed consisting of $25$ identical Pycom IoT devices and a USRP B$210$ receiver, operating at a center frequency of $915$MHz, used for recording the received signals sampled at $1$MS/s. Recorded data in the form of both the time-domain I/Q samples and FFT samples are stored into binary files in compliance with SigMF~\cite{hilburn2018sigmf} by creating, for each binary file, a metafile written in plain-text JSON to include recording information such as sampling rate, time and day of recording, and carrier frequency, among other parameters. 

This dataset covers multiple experimental setup scenarios, which are summarized in Table \ref{summary} and can be downloaded at
\href{https://research.engr.oregonstate.edu/hamdaoui/datasets/}{\color{blue}{http://research.engr.oregonstate.edu/hamdaoui/datasets}}. 

The experimental scenarios are briefly described next, with more details provided in~\cite{elmaghbub2021lora}.
 \begin{itemize}
         \item {\bf Setup 1---Different Days Indoor Scenario:} Indoor setup with data collected over $5$ consecutive days. For each day, $10$ transmissions were captured from each of the 25 transmitters, with each transmission having a duration of $20$s. All transmissions took place $1$ minute apart from one another. 
         \item {\bf Setup 2---Different Days Outdoor Scenario:} Outdoor setup with data collected over $5$ consecutive days. For each day, $10$ transmissions were captured from each of the $25$ transmitters, with each transmission having a duration of $20$s. All transmissions took place $1$ minute apart from one another. 
         \item {\bf Setup 3---Different Days Wired Scenario:} Wired setup with data collected over $5$ consecutive days. For each day, $10$ transmissions were captured from each of the $25$ transmitters, with each transmission having a duration of $20$s. All transmissions are $1$ minute apart from one another. 
         \item {\bf Setup 4---Different Distances Scenario:} Outdoor setup with data collected from $4$ different distances: $5$m, $10$m, $15$m, and $20$m away from the receiver. For each distance, one transmission was captured from each of the $25$ transmitters, with each transmission having a duration of $20$s. All transmissions are $1$ minute apart from one another. 
         \item {\bf Setup 5---Different Configurations Scenario:} Indoor setup with data collected from $4$ different LoRa configurations. For each configuration, one transmission was captured from each of the $25$ transmitters, with each transmission having a duration of $20$s. All transmissions are $1$ minute apart from one another. 
         \item {\bf Setup 6---Different Locations Scenario:} This data has been collected in $3$ different locations: room, outdoor, and office environments. At each location, one transmission was captured from each of the $25$ transmitters, with each transmission having a duration of $20$s. All transmissions are $1$ minute apart from one another.
         \item {\bf Setup 7---Different Receivers Scenario:} Indoor setup with data collected using $2$ different receivers. For each receiver, one transmission was captured from each of the $25$ transmitters, with each  having a duration of $20$s. All transmissions are $1$ minute apart from one another. 
\end{itemize}

The rest of the paper is organized as follows. Section \ref{sec:testbed} describes the testbed components. Sections~\ref{experiment} and~\ref{dataset} describe the different experimental setups and the dataset, respectively. Section~\ref{use-case} presents a use case for the dataset. The challenges, limitations, and new opportunities are discussed in Section \ref{challenges} and the paper is concluded in Section \ref{Conclusion}.

\label{intro}

\section{Testbed}
\label{sec:testbed}
In this section, we describe the hardware, software, and protocol components used in building our testbed.
\subsection{Hardware Description}
\begin{figure}
    \centering
    \includegraphics[width=1\columnwidth, height=0.57\columnwidth]{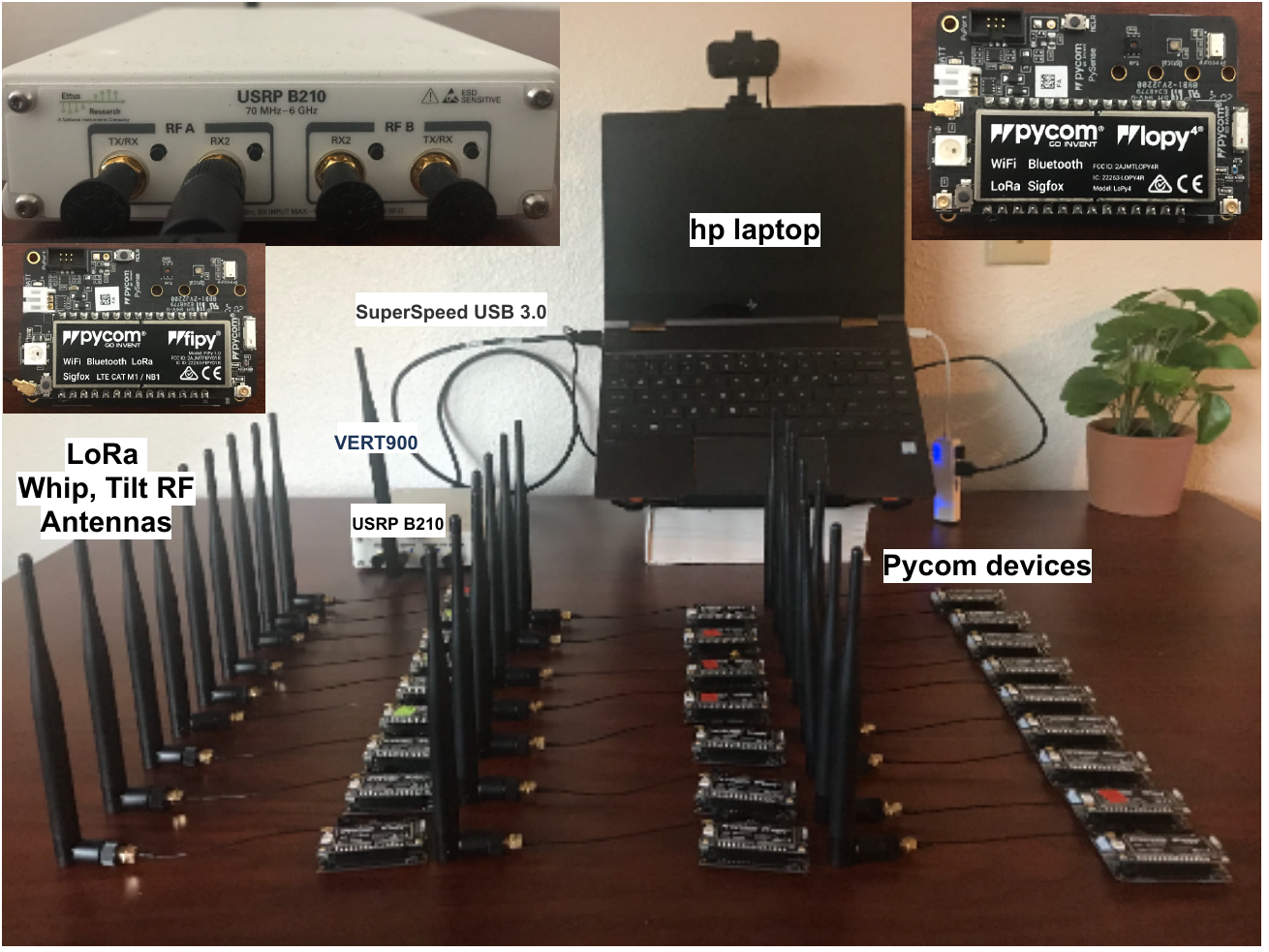}
    \caption{IoT Testbed: $25$ Pycom devices and USRP B$210$.}
    \label{fig:dataset}
\end{figure}

\begin{figure*}
    \centering
    \includegraphics[width=1.7\columnwidth, height=0.5\columnwidth]{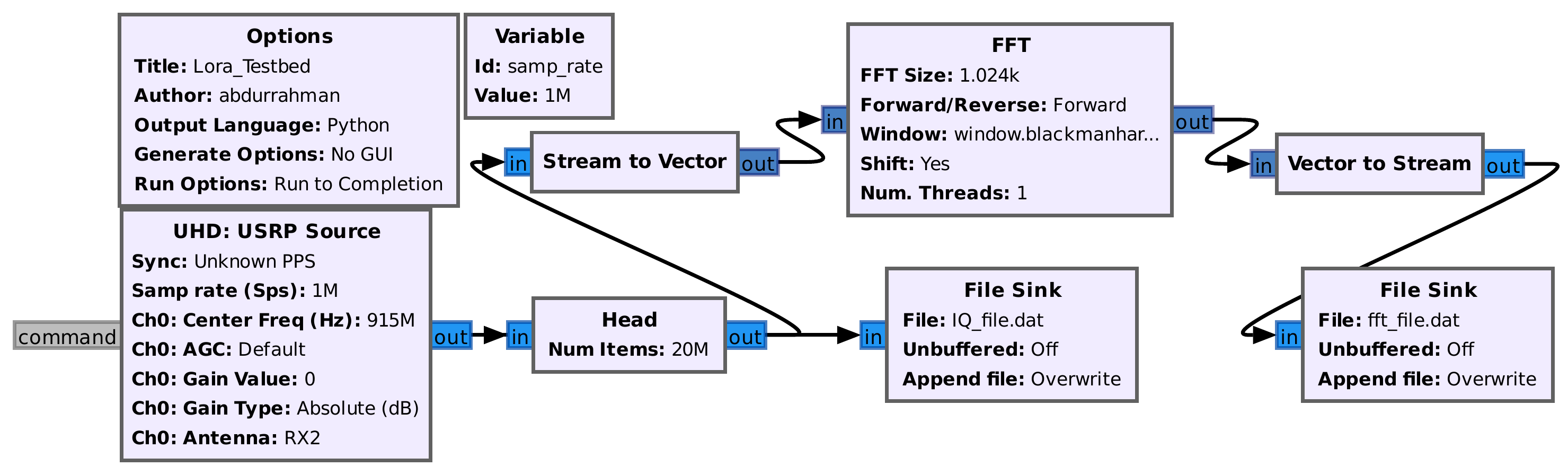}
    \caption{The flowgraph of our data collection.}
    \label{fig:flow}
\end{figure*}
Our IoT testbed, shown in Fig.~\ref{fig:dataset}, consists of $25$ Pycom devices with Semtech SX$1276$ LoRa transceivers, $25$ Pycom sensor shields, and an Ettus USRP (Universal Software Radio Peripheral) B$210$ for data sampling, which was configured with  a center frequency of $915$MHz and a sample rate of $1$MS/s. Our collection of Pycom devices is made up of $23$ Lopy4 boards and $2$ Fipy boards, which are MicroPython-enabled development boards with multiple network capabilities: LoRa, Sigfox, WiFi, Bluetooth, and NB-IoT. The sensor shield collection comprises $22$ PySense boards equipped with sensing capability, $2$ PyScan boards equipped with RFID scanning capability, and $1$ PyTrack board equipped with GPS/tracking capability. We used lipo batteries to power the devices. Each Pycom device was connected to a dedicated LoRa antenna and configured to transmit LoRa transmissions at the $915$MHz US band adapting the following configuration: Raw-LoRa mode, $125$KHz bandwidth, a spreading factor (SF) of $7$, a preamble of $8$, a TX power of $20$dBm, and a coding rate of $4/5$. 

\subsection{Software Description}
\subsubsection{Transmission Subsystem}
We programmed and configured our Pycom boards using MicroPython \cite{tollervey2017programming}, which is an efficient implementation of Python3 that is composed of a subset of standard Python libraries and optimized to run on microcontrollers and constrained environments. Also, we used Pymakr plugin as a REPL console that connects to Pycom boards to run codes or upload files. 

\subsubsection{Reception Subsystem}
We used the GNURadio software \cite{valerio2008open}, a real-time signal processing graphical tool, to set up and configure the USRP receiver to capture LoRa transmissions, plot their time and spectrum domains, implement some preprocessing techniques and store the samples into their files. Fig.~\ref{fig:flow} shows the general flowgraph used for data acquisition. 

\subsection{LoRa Protocol Description}
We transmitted/captured LoRa modulation signals, a proprietary physical layer implementation that employs Chirp Spread Spectrum (CSS) in the sub-GHz ISM band and trades data rate for coverage range, power consumption, or link robustness. LoRa does so by providing a tunable parameter, called a spreading factor (SF), which varies from $7$ to $12$ and determines the sequence length of an encoded symbol within a fixed bandwidth. A higher spreading factor means longer ranges with lower data rates. Unlike other spread spectrum techniques, the chirp-based modulation allows LoRa to maintain the same coding gain and immunity to noise and interference while meeting the low-cost, low-power consumption requirements. A LoRa modulator generates both raw chirp signals with fixed amplitude and continuously varying frequency with constant rate and a set of modulated chirps that are cyclically time-shifted raw-chirps where the initial frequency determines the content of the chirp symbol.

\section{Experimental Setups}
\label{experiment}
\begin{figure}
    \centering
    \includegraphics[width=0.9\columnwidth]{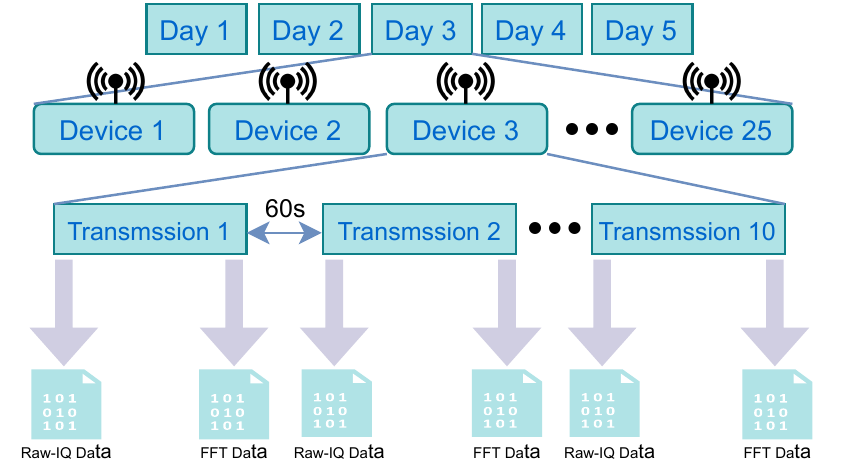}
    \caption{Setup diagram: setups 1, 2 and 3.}
    \label{fig:flow1}
\end{figure}

We use our testbed described in Section~\ref{sec:testbed} to create and collect large-scale, comprehensive LoRa RF fingerprinting dataset for multiple experimental scenarios that are specifically designed to allow intensive and comprehensive performance evaluation of various deep learning-based wireless networks techniques, with a special focus on RF/device fingerprinting. We chose a bandwidth of $125$KHz for all LoRa transmissions. However, in \cite{elmaghbub2020widescan}, we demonstrated that considering out of band spectrum enhances the performance of the model, and hence we provide a total bandwidth of $1$MHz that covers the in-band as well as an adjacent out of band spectrum of LoRa transmissions for all setups. In the remainder of this section, we present each of the seven considered experimental setups. Throughout this section, we will be referring to Fig.~\ref{fig:flow1} for experimental setups 1, 2, and 3, and for Fig.~\ref{fig:flow2} for experimental setups 4, 5, 6 and 7. Table~\ref{summary} summarizes these 7 setups.

\subsection{Setup 1: Different Days Indoor Scenario}
In order to enable performance evaluation while masking the impact of the outside environment, we created an indoor setup, ran experiments, and collected datasets for this setup. These indoor experiments were carried out in a typical occupied room environment over $5$ consecutive days. All devices transmitted the same message from the same location, $5$m away from the receiver so that all devices experience similar channel conditions. As shown in Fig.~\ref{fig:flow1}, for each day, each transmitter generated $10$ transmissions, each of $20$s duration, all spaced apart by $1$ minute. Hence, we collected about $200$M complex-valued samples from each device per day. We used GNURadio packages to store the sampled raw-I/Q values and their corresponding FFT-Based representation into binary files as depicted in Fig.~\ref{fig:flow1}.

\subsection{Setup 2: Different Days Outdoor Scenario}
In order to allow for performance evaluation while considering the impact of outdoor wireless channel impairments, we carried out the experiments in an outdoor environment at nighttime. Here again, all devices transmitted the same message from the exact location, situated $5$m away from the receiver, so that all devices experience similar channel conditions. Like in the indoor setup case and as shown in Fig.~\ref{fig:flow1}, for five consecutive days, each transmitter generated $10$ transmissions per day, each of $20$s duration, all spaced $1$ minute apart from one another. This resulted in about $200$M complex-valued samples per device per day. We ran this experiment over $5$ consecutive days and provided 5-day worth of data to study the robustness of deep learning models when trained on data collected on one day but tested on data captured on a different day.  We used GNURadio packages to store the sampled raw-I/Q values and their corresponding FFT-based representation into binary files as depicted in Fig.~\ref{fig:flow1}.

\subsection{Setup 3: Different Days Wired Scenario} The wireless channel has a notable impact on the performance of deep learning models and presents its unique challenges \cite{al2020exposing}. Hence, to assess how well these models perform in the absence of the wireless channel's impact, we created a wired setup where the Pycom boards are directly connected to the USRP via an SMA cable and $30$dB attenuator. Similar to the Indoor and Outdoor experiments, we ran this experiment over $5$ consecutive days. For every day, each device transmitted $10$ bursts each of $20$s duration. Therefore, the total amount of collected data is $200$M complex-valued samples per device per day. We again used GNURadio packages to store the sampled raw-I/Q values and their corresponding FFT-based representation into binary files as depicted in Fig.~\ref{fig:flow1}.

\subsection{Setup 4: Different Distances Scenario}
\begin{figure}
    \centering
    \includegraphics[width=1\columnwidth]{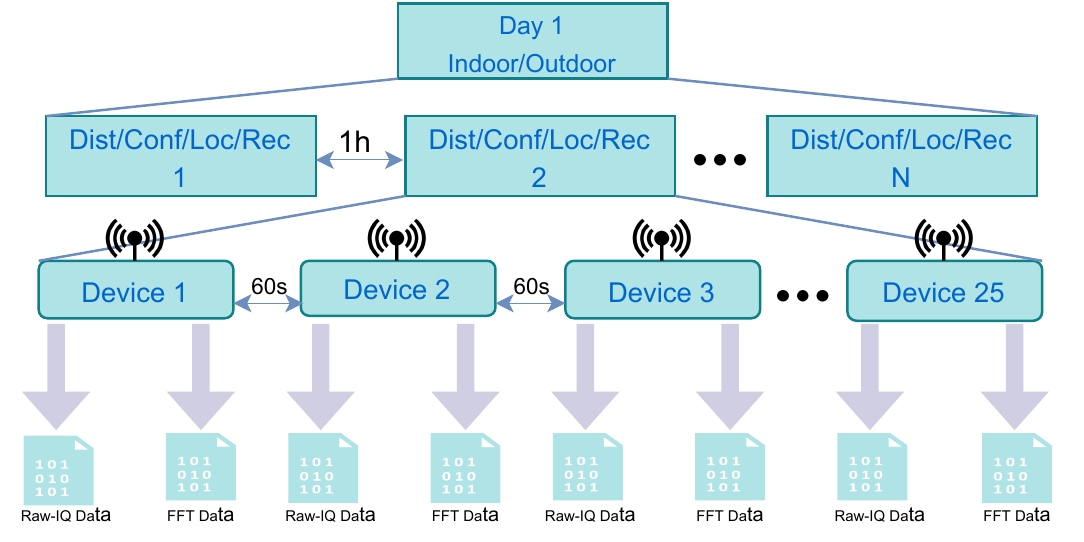}
    \caption{Setup diagram: setups 4, 5, 6, and 7.}
    \label{fig:flow2}
\end{figure}

Some end-devices constantly change their positions, so it is critical to explore the impact of distance on the performance of classifiers and see whether or not a classifier would still recognize a device when it moves to a position that is different from the one used for training. This experiment was carried out in a typical outdoor environment in a sunny day. We considered four different distances, $5$m, $10$m, $15$m, and $20$m, and for each distance, we collected $1$ transmission of $20$s for each of the $25$ devices. We kept the receiver at the same location for all the transmissions while locating the transmitters at $4$ different distances away from the receiver base ($5$m, $10$m, $15$m, and $20$m). The transmissions were captured consecutively in time with only $60$s apart from one another. Each transmitter generated $4$ transmissions each of $20$s duration, resulting in $80$M complex-valued samples from each device. We again used GNURadio packages to store the sampled raw-I/Q values and corresponding FFT-based samples into binary files as depicted in Fig.~\ref{fig:flow2}.

\subsection{Setup 5: Different Configurations Scenario}
\begin{figure}
    \centering
    \includegraphics[width=1\columnwidth,height=0.53\columnwidth]{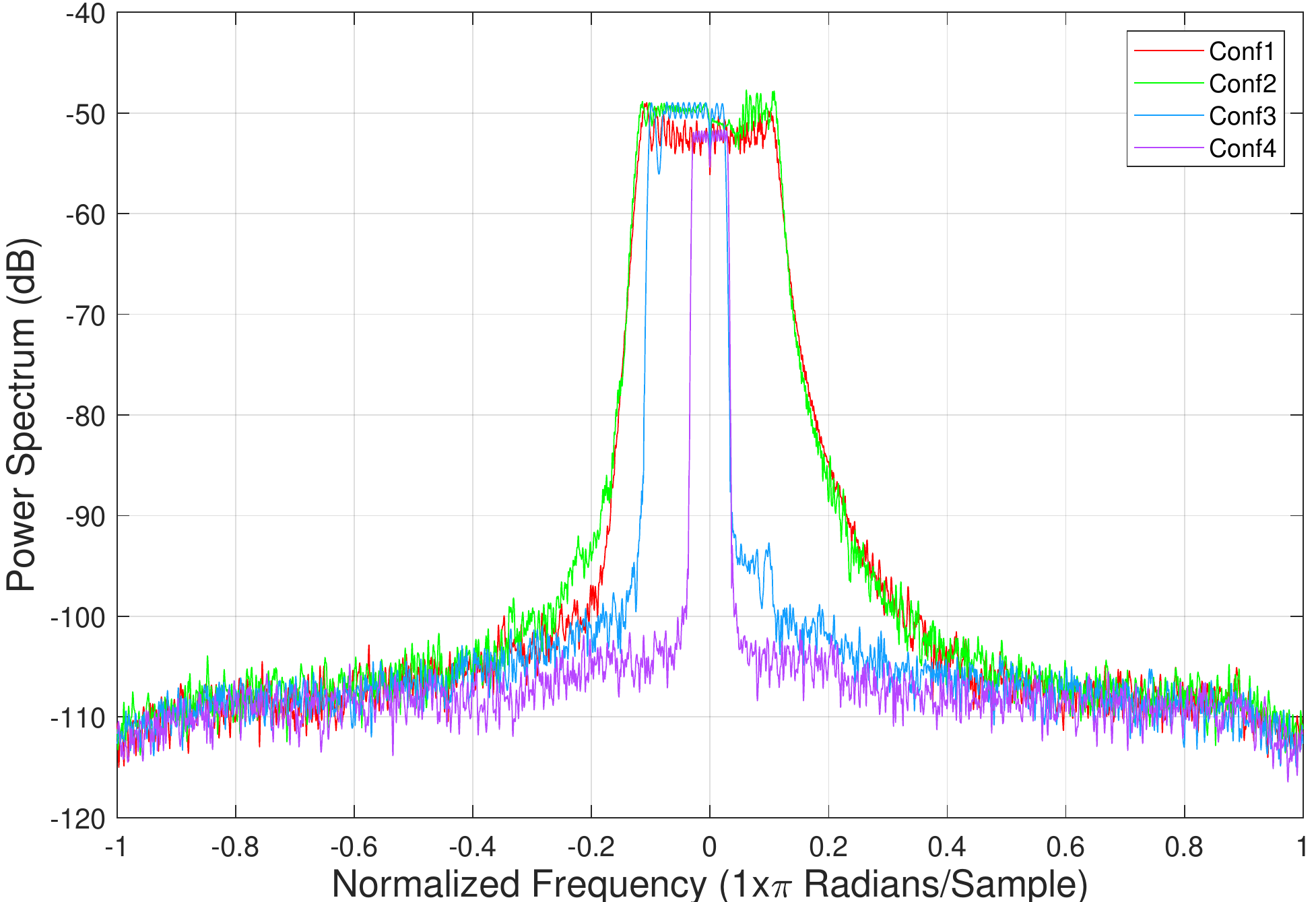}
    \caption{Spectrum of the 4 LoRa configurations from Setup 5}
    \label{fig:spec4}
\end{figure}

LoRaWAN uses the Adaptive Data Rate (ADR) mechanism to optimize the data rates, air-time, and energy consumption in the network to accommodate the varying RF conditions. This mechanism allows the network server to inform end devices to adjust their power consumption and data rate as needed. This is achievable by controlling the following parameters at end devices: spreading factor, bandwidth, and power consumption. Changing the spreading factor, for example, in LoRa modulation results in a change in the data rate, receiver sensitivity, time in the air, and power consumption. Fig.~\ref{fig:spec4} shows the frequency spectrum of a snapshot of the four LoRa configurations that we included in our dataset. Ideally, a classification model (e.g., a deep learning fingerprinting model) should identify a device even if it changes its configuration; i.e., models that are trained using one configuration but tested on a different configuration should still perform well. Therefore, in order to enable the assessment of how agnostic these models are to protocol configuration, we captured transmissions using $4$ different configurations, as presented in Table~\ref{con}. For this, we collected a single LoRa transmission of $20$s from each device for each configuration in an indoor setup with $5$m as the distance between the receiver and transmitters. Like other setups, for this setup, we used GNURadio packages to store the sampled raw-I/Q values and their corresponding FFT-based representation into binary files as depicted in Fig.~\ref{fig:flow2}.
\subsection{Setup 6: Different Locations Scenario}
Another practical scenario we consider here aims at allowing deep learning models to be trained on data captured in one location but tested on data collected in another location. For this, we captured LoRa transmissions in three different location deployments, room environment, office environment, and outdoor environment, all taken on the same day. Here, we kept the distance between the receiver and transmitters (i.e., $5$m) and the LoRa configuration the same. We captured a single transmission of $20$s from each device at each location with a $60$s period between the devices, resulting in $60$M complex samples from each device. We again used GNURadio packages to store the sampled raw-I/Q values and their corresponding FFT-based representation into binary files as depicted in Fig.~\ref{fig:flow2}.
\begin{table}
\scriptsize
\begin{tabular}{|l|c|c|l|c|c|}
\hline
\textbf{Configurations} & \multicolumn{1}{l|}{{ \textbf{\begin{tabular}[c]{@{}l@{}}Spread\\ Factor\end{tabular}}}} & \multicolumn{1}{l|}{{ \textbf{Bandwidth}}} & \textbf{Bit Rate} & \multicolumn{1}{l|}{\textbf{\begin{tabular}[c]{@{}l@{}}Tx \\ Power\end{tabular}}} & \multicolumn{1}{l|}{\textbf{\begin{tabular}[c]{@{}l@{}}Coding\\ Rate\end{tabular}}} \\ \hline
Configuration 1 & { 7}                                                                                     & {\color[HTML]{333333} 125KHz}                                  & 5470 bps          & 20dBm                                                                              & 4/5                                                                                 \\ \hline
Configuration 2 & 8                                                                                                            & 125KHz                                                         & 3125 bps          & 20dBm                                                                              & 4/5                                                                                 \\ \hline
Configuration 3 & 11                                                                                                           & 125KHz                                                         & 537 bps           & 20dBm                                                                              & 4/5                                                                                 \\ \hline
Configuration 4 & 12                                                                                                           & 125KHz                                                         & 293 bps           & 20dBm                                                                              & 4/5                                                                                 \\ \hline
\end{tabular}

\caption{LoRa Configurations}
\label{con}

\end{table}

\subsection{Setup 7: Different Receivers Scenario}
Like transmitters, receivers also suffer from hardware impairments due to hardware imperfection. Therefore, deep learning models trained using data collected by one receiver but tested using data collected by a different receiver may not perform well due to the possible additional impairments caused by the receiver's reception.
To allow researchers to study the impact of such a change in the receiving device, we provided a dataset for the 25 Pycom transmitting devices, collected using two different USRP B$210$ receivers. In this experiment,  we considered an indoor setup where the transmitters (the 25 Pycom devices) were located $5$m away from the receiver. For each receiver, we captured a single transmission of $20$s from each of the 25 Pycom transmitters, yielding a total of $40$M samples for each device. Like other setups, for this setup, we used GNURadio packages to store the sampled raw-I/Q values and their corresponding FFT-based representation into binary files as depicted in Fig.~\ref{fig:flow2}.

\section{Dataset Description}
\label{dataset}
\begin{figure*}[t!]
    \centering
    \includegraphics[height=0.83\columnwidth, width=2\columnwidth]{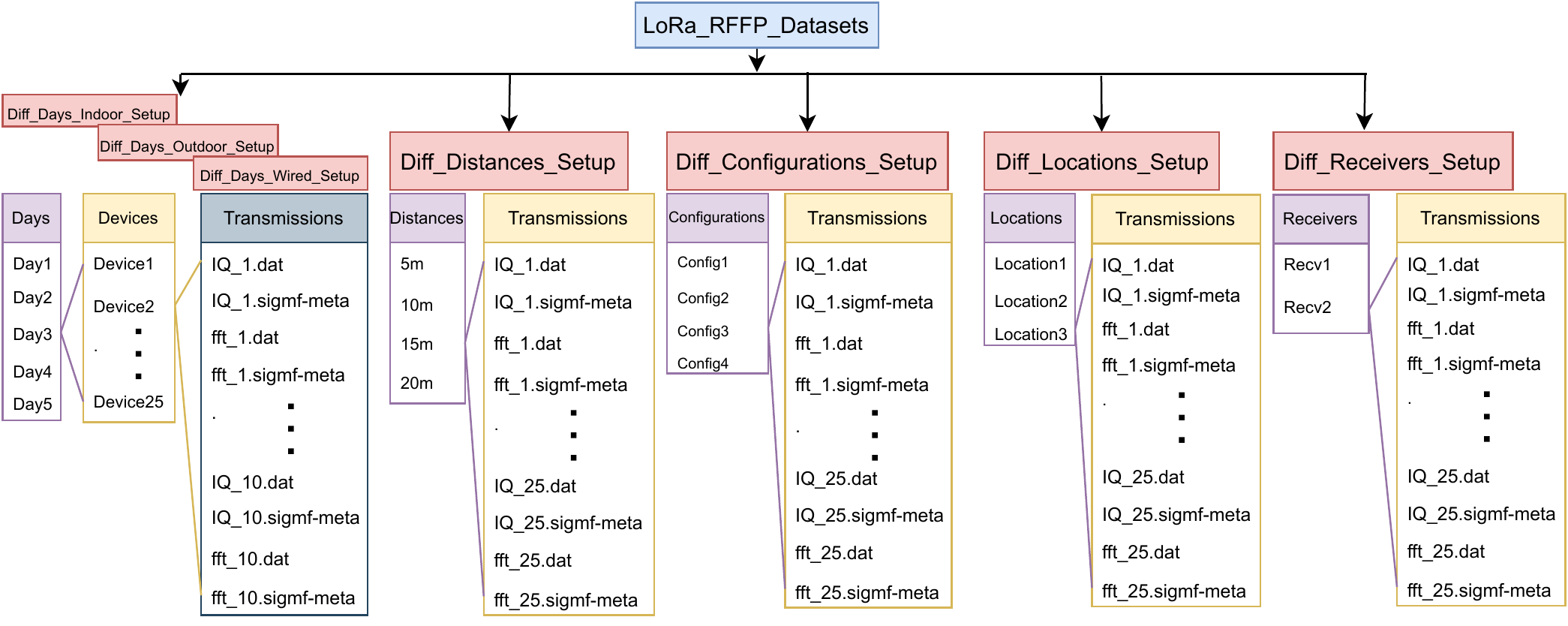}
    \caption{Online link/file organization of the datasets.  Note that Diff\_Days\_Indoor\_Setup, Diff\_Days\_Outdoor\_Setup, and Diff\_Days\_Wired\_Setup directories have the same file system structure.}
    \label{fig:orgnization}
\end{figure*}
Our dataset contains LoRa transmissions of $25$ devices and a total of $43$ transmissions on average for each device. An Ettus USRP B$210$ is used to record each transmission, operating at a center frequency of $915$MHz with a sampling rate of $1$MS/s. Each recording consists, on average, of $20$M I/Q samples. For each recording, we stored the time-domain I/Q samples and FFT-based samples into binary files. The binary files are encoded with Float$32$, and the complex-valued sampled are interleaved where the I-values are in the odd indices, and the Q-values are in the even ones. In order to be SigMF (Signal Metadata Format)~\cite{hilburn2018sigmf} compliant, we created a metadata file written in plain-text JSON adapting SigMF for each binary file to describe the essential information about the collected samples, the system that generated them, and the features of the signal itself. In our case, we stored in the metadata files information regarding $(i)$ the sampling rate, $(ii)$ time and day of recording, and $(iii)$ the carrier frequency, among others. 


More details and use cases of these LoRa datasets can be found in~\cite{elmaghbub2021lora}. The datasets can be downloaded from NetSTAR Lab at 
\href{https://research.engr.oregonstate.edu/hamdaoui/datasets/}{\color{blue}{http://research.engr.oregonstate.edu/hamdaoui/datasets}}.
Users of the datasets may refer to Fig.~\ref{fig:orgnization} for help with the file organization and notation.  Specifically, these are:


\begin{itemize}
\item {\bf Diff\_Days\_Indoor\_Setup}, {\bf Diff\_Days\_Outdoor\_Setup}, and {\bf Diff\_Days\_Wired\_Setup} directories (correspond to Setups 1, 2 and 3), each having 5 subdirectories, one for each day. Each day subdirectory has $25$ subdirectories, one for each device. Each device subdirectory has $40$ files (for $10$ transmissions): $10$ I/Q data files plus their corresponding $10$ metadata files and $10$ fft data files plus their corresponding 10 metadata files.
%

\item {\bf Diff\_Distances\_Setup} directory (corresponds to Setup 4), having $4$ subdirectories representing the four distances. Each subdirectory includes 100 files: $25$ I/Q data files plus their corresponding $25$ metadata files and $25$ fft data files plus their corresponding $25$ metadata files.


\item {\bf Diff\_Configurations\_Setup} directory (corresponds to Setup 5), having $4$ subdirectories representing the four configurations. 
Each of these 4 subdirectories includes 100 files: $25$ I/Q data files plus their corresponding $25$ metadata files and $25$ fft data files plus their corresponding $25$ metadata files.

\item {\bf Diff\_Locations\_Setup} directory (corresponds to Setup 6), having $3$ subdirectories representing the three locations. 
Each subdirectory includes 100 files: $25$ I/Q data files plus their corresponding $25$ metadata files and $25$ fft data files plus their corresponding $25$ metadata files.

\item {\bf Diff\_Receivers\_Setup} directory (corresponds to Setup 7), having $2$ subdirectories representing the two receivers. 
Each subdirectory includes 100 files: $25$ I/Q data files plus their corresponding $25$ metadata files and $25$ fft data files plus their corresponding $25$ metadata files.
\end{itemize}

\begin{figure}[h!]
    \centering
    \includegraphics[width=1\columnwidth, height=0.75\columnwidth]{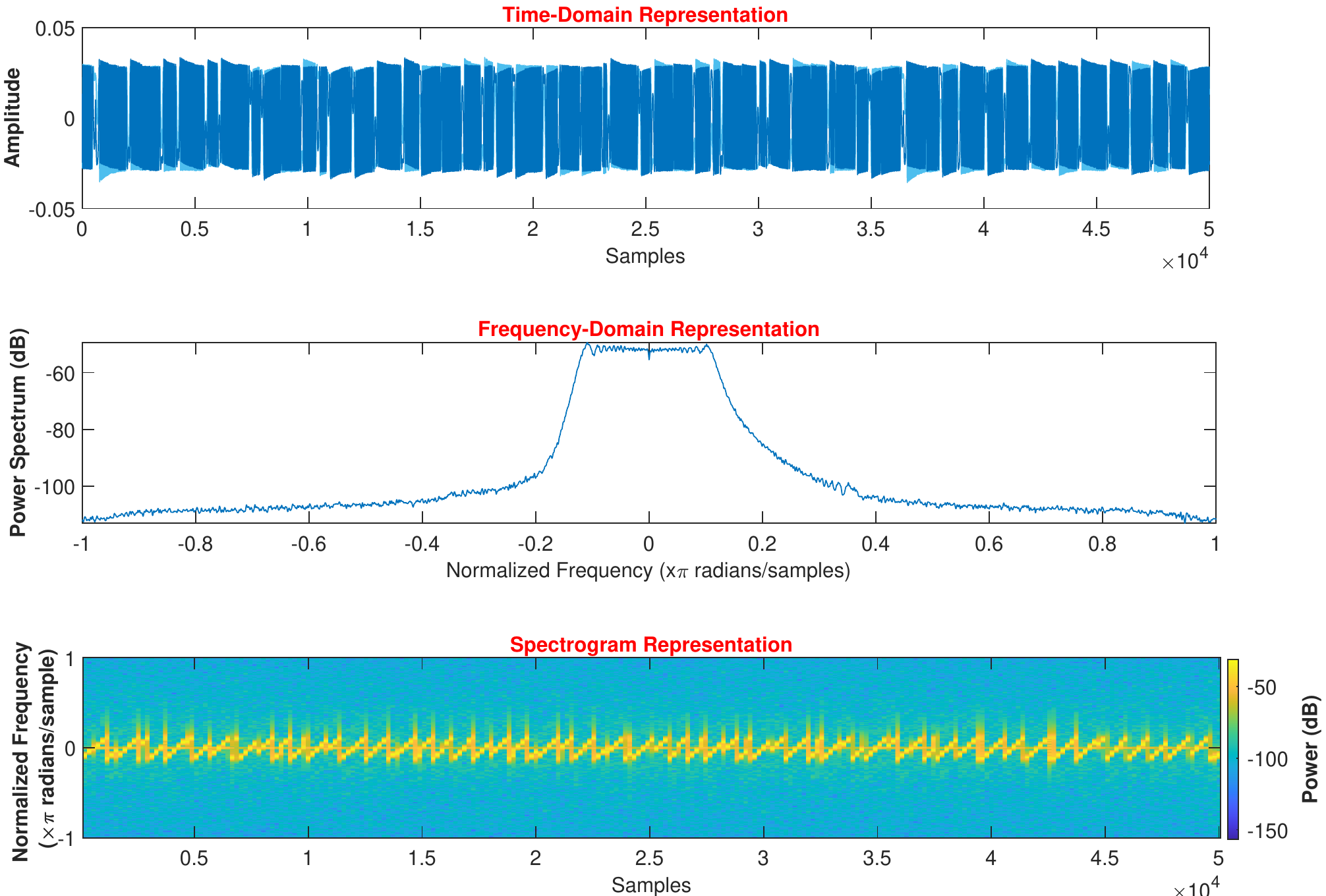}
    \caption{Top: I/Q time-domain representation, Middle: Frequency spectrum representation, Bottom: Spectrogram representation from Device 1, Indoor Setup.}
    \label{fig:vis}
\end{figure}

To visualize the captured LoRa signals, we \hl{plotted}, in Fig.~\ref{fig:vis}, the time-domain, frequency spectrum, and spectrogram of a LoRa transmission captured in an Indoor environment. The top figure in Fig.~\ref{fig:vis} represents the in-phase (I) and quadrature (Q) components of the time-domain signal from device 1 in an indoor setup, while the figure in the middle shows its frequency spectrum and the figure in the bottom shows the upward chirps in the spectrogram of the same LoRa signal.

\section{Use Case: LoRa Device Fingerprinting}
\label{use-case}
LoRaWAN~\cite{sundaram2019survey}, a low-power wide-area network technology, has emerged as the de facto standard for connecting thousands of million devices serving many IoT ecosystems and applications. Its ability to transmit over long distances with an optimized low-power consumption, leading to a long battery life of up to 10x that of cellular M2M technologies, makes it a good fit for next-generation IoT connections. LoRaWAN defines two layers of security: network-level security that deals with LoRa device authenticity and message integrity, and application-level security that provides end-to-end data encryption to ensure private device-to-server communication~\cite{basu2020security}. The functionality of these layers relies on generating and storing session keys (AppKey and NwkKey) in a secured manner. Some IDs and keys, such as DevEUI and AppKey, can be hard-coded on devices' tags or software, and thus simple but common human errors, such as failing in removing the tags or replacing the source code before deployment, can expose the network to security risks~\cite{lora_sec}. 

It is essential to ensure that conventional high-layer security mechanisms are complemented with unclonable, physical-layer security solutions so as to increase the security robustness of such systems~\cite{5751298}.  One technique that has been considered for providing physical-layer security is to exploit wave-level distortions in the received RF signals that are generated by hardware impairments to provide unique signatures (aka fingerprints) of the devices, which can serve as device identifiers (e.g.,~\cite{sankhe2019oracle, hamdaoui2020deep}). The uniqueness of RF fingerprints comes from the collective random deviations of a tremendous number of RF analog components from their ideal values during the manufacturing process. Hence, we can confidently claim that there are no two analog circuits with identical collective deviations profiles, and therefore, there are no two circuits with the same RF fingerprint. IQ imbalance, DC offset, phase noise, and power amplifier nonlinearity are among other transceiver hardware impairments that manifest in uniquely distinguishing features that increase devices' separability in RF fingerprinting techniques~\cite{elmaghbub2020leveraging}.

Until recently, the feature selection process for RF fingerprinting techniques has been done manually, which requires expert knowledge and many trials and error iterations to find the features that guarantee the best performance. Nevertheless, in most cases, they end up with protocol-specific or vendor-specific solutions. Following the unprecedented achievements of deep learning models in computer vision and natural language processing in recent years, researchers have shown that the function approximation power of deep learning models can be leveraged to better improve the classification performance of RF fingerprinting techniques and other RF-domain challenges. These results led to a rush in research activities in the area of machine learning for RF systems from both industry and academia, creating an urgent need for comprehensive RF datasets that can be used for validating their proposed models.

\section{Challenges, Limitations and Opportunities}
\label{challenges}
We next describe some limitations and challenges faced when creating RF fingerprint datasets.
\begin{itemize}
\item {\bf Same channel condition.} The time between a given device's first and last transmissions may not be short enough for all transmissions to be assumed to be taken under the same condition. This makes it challenging to study the performance of a learning model with training and testing being done under the same channel condition. This is because all 25 devices must first be sampled for their first transmissions before moving to the subsequent transmission. Even if one tries to collect all 10 transmissions sequentially for each device to minimize this timing effect between transmissions, we run into the problem of increasing the experiment time between devices.


\item {\bf Same power levels.} Maintaining the same power level for all devices is important so that to mask the power impact. In our experiments, to mitigate this issue, we start with a full-charged battery every time, though this still cannot guarantee that all have the same power level.

\item {\bf Concurrent transmissions.} One interesting scenario but difficult to realize is the ability to collect data from multiple devices while all transmitting concurrently, as it often occurs in random access procedures for wireless base stations. Excelling on this scenario would open the door for incorporating deep learning-based fingerprinting into next-generation cellular networks. 
    
\item {\bf Devices at scale.} Increasing the number of transmitters in our testbed is another to-do-item that would add more credibility to the evaluation and allow us to assess the scalability performance of the proposed models.
    
\item {\bf Beyond RF fingerprinting.} Leveraging the multi-bearer capability of our testbed, one can use the testbed to collect datasets for modulation recognition using other technologies like LoRaWAN, SigFox, Bluetooth, WiFi, and NB-IoT technologies. Another use is for creating datasets for studying indoor device positioning problems.  
\end{itemize}

\section{Conclusion}
\label{Conclusion}
This paper presents a comprehensive LoRa RF fingerprint datasets for multiple experimental scenarios specifically designed to allow thorough performance assessment of deep learning-based wireless networks techniques, such as RF/device fingerprinting. This dataset is made available to the research community to serve as a benchmark for testing RF classification and fingerprinting techniques.


\section{Acknowledgment}
\label{ack}
This research is supported in part by Intel/NSF MLWiNS Award No. 2003273. We would like to thank Intel researchers, Dr. Kathiravetpillai Sivanesan, Dr. Lily Yang, and Dr. Richard Dorrance, for their constructive feedback. 

\vspace{12pt}
\color{black}
\bibliographystyle{IEEEtran}
\bibliography{IEEEexample}
\end{document}